\documentstyle[prl,aps,twocolumn,epsf]{revtex}

\begin{document}
\draft
\wideabs{
\title{Zeolite-dye  micro lasers}
\author{U. Vietze, O. Krau\ss, F. Laeri
\thanks{\quad Corresponding author:  Franco Laeri,
Darmstadt University of Technology, Institute of Applied Physics,
D-64289 Darmstadt (Germany); Tel.:  +49-6151-16 5495, Fax.:
+49-6151-16 3022, Email:  franco.laeri@physik.tu-darmstadt.de }}
\address{Darmstadt University of Technology, D-64289 Darmstadt}
\author{G. Ihlein, F. Sch\"uth}
\address{University of Frankfurt, D-60439 Frankfurt}
\author{B. Limburg, M. Abraham}
\address{IMM, D-55129 Mainz}
\date{\today}
\maketitle
\begin{abstract}
We present a new class of micro lasers based on nanoporous
molecular sieve host-guest systems.  Organic dye guest molecules
of 1-Ethyl-4-(4-(p-Dimethylaminophenyl)-1,3-butadienyl)-pyridinium
Perchlorat were inserted into the 0.73-nm-wide channel pores of a
zeolite AlPO$_4$-5 host. The zeolitic micro crystal compounds
where hydrothermally synthesized according to a particular
host-guest chemical process. The dye molecules are found not only
to be aligned along the host channel axis, but to be oriented as
well. Single mode laser emission at 687~nm was obtained from a
whispering gallery mode oscillating in a 8-$\mu$m-diameter
monolithic micro resonator, in which the field is confined by
total internal reflection at the natural hexagonal boundaries
inside the zeolitic microcrystals.
\end{abstract}
}

\narrowtext
Structures of optical wavelength scale such as micro resonators and
micro lasers are presently attracting considerable attention in
quantum optics, laser physics, materials science, and device
technology.  The reduction of at least one lateral dimension of a
laser structure down to the order of one wavelength modifies  the
mode density of the optical field inside the laser resonator which
in turns modifies the spontaneous emission rate \cite{YOK95}. As a
result, the light generation efficiency in wavelength--scale micro
lasers is expected to increase. In addition, the cavity quantum
electrodynamic effects that are predicted to appear in this regime
can be exploited to accomplish novel functions \cite{RAR96,DUC96}.

From an experimental point of view, reducing the size of a laser
necessitates that careful attention be given to the scaling laws of
the two basic parameters gain and losses. On one hand the gain of the
light propagating in the laser structure scales in general with the
propagation length, as do the scattering losses in the medium. On the
other hand, the coupling losses from the inside to the outside of the
laser structure do not depend on geometrical factors. Consequently,
even for laser media with very low scattering losses, scaling down to
the size of a few wavelengths becomes difficult because the reduced
gain can no longer compensate for the coupling losses. Up to now,
laser size reduction down to a few ten micrometers has been
sucessfully carried out for media with high gain, such as
semiconductors  (vertical cavity surface emitting lasers, microdisk
\cite{RAR96,GMA98}, or spherical lasers \cite{NAG97}), and sperical
liquid \cite{TZE84}, or polymer dissolved dyes \cite{KUW92,CHA96}, as
well as solid state spheres \cite{SAN96}. Thus, realizing wavelength
scale lasers based on alternative materials, geometries, and emission
wavelengths is still a challenging experimental task. As a reward, in
addition to a gain in their basic understanding, one can imagine many
practical applications for microlasers, e.g.~as efficient
luminophores, or as bright lasing pixels for displays \cite{LAW97}.

In this letter, we report the first realization of a new class of
solid state micro lasers based on a host-guest composite material. In
contrast to conventional gain materials, in which  the active centers
are not systematically ordered on the microscopic level, the regular
crystallographic matrix of cavities in a zeolite crystal maintains the
laser active organic guest molecules aligned and oriented. In fact,
the aluminosilicate structure of zeolitic crystals surrounds regularly
arranged cavities or channels with a size of up to ca.\ $1\times
10^{-9}$~m \cite{SMI88,ZOL60}. It is this framework of
crystallographically arranged nanometer cavities that has stimulated
physicists, chemists, and materials scientists to investigate
guest-host modifications of the host material on the nanometer scale,
with the objective of tailoring the properties of the composite on the
macroscopic scale.  For example, Bogomolov et al.\ \cite{BOG76}
synthesized nanocomposites of semiconductor based clusters in zeolites
and superconducting quantum wires \cite{BOG73}, whereas Cox
et~al.~\cite{COX88} presented a concept for the modification of the
nonlinear optical response of molecular sieve materials.  Further
applications based on the inclusion concept were proposed, cf.\ e.g.\
\cite{HER89,OZI89,DOU90}.  For optical functions, the inclusion ansatz
was first successively embodied by Cox et al.\ \cite{COX90}.  The
group of Caro and Marlow \cite{MAR92} also transformed various AlPO
and SAPO molecular sieves into a nonlinear material for optical
frequency conversion by inclusion of the noncentrosymmetric organic
molecule $p$-nitroaniline, and Bredol et~al.~\cite{BRE91} included
efficient luminophores in zeolites.

\begin{figure}[t]
\begin{center}
\epsfxsize=80mm \epsffile{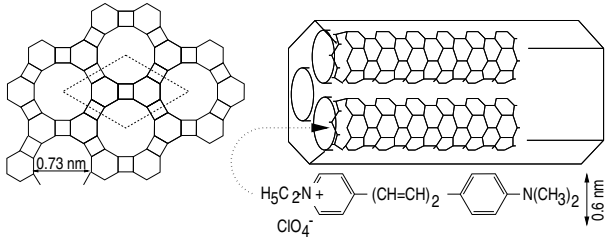}
\end{center}
\caption{ The zeolite AlPO$_4$-5, its one dimensional channel
framework, and the inserted dye molecule
1-Ethyl-4-(4-(p-Dimethylaminophenyl)-1,3-butadienyl)-pyri\-dinium
Perchlorat (Pyridine~2). } \label{DyeMol}
\end{figure}

The framework of the hexagonal ($\rm P \frac{6}{\rm m}cc$)
AlPO$_4$-5 crystals  discussed here  consists of a one-dimensional
system of pores (channels) 0.73~nm wide. Guided mainly by
geometrical considerations we selected the dye molecule
1-Ethyl-4-(4-(p-Di\-methylaminophenyl)-1,3-butadienyl)-pyridinium
Perchlorat (Pyridine~2) \cite{BRA94} as luminescent activator
center. With a width of ca.~0.6~nm (calculated with the code
SHELXTL) it fits into the one-dimensional channel framework of the
zeolite AlPO$_4$-5 \cite{MEI87}; cf.~fig.~\ref{DyeMol}.

\begin{figure}
\epsfxsize=178mm \epsffile{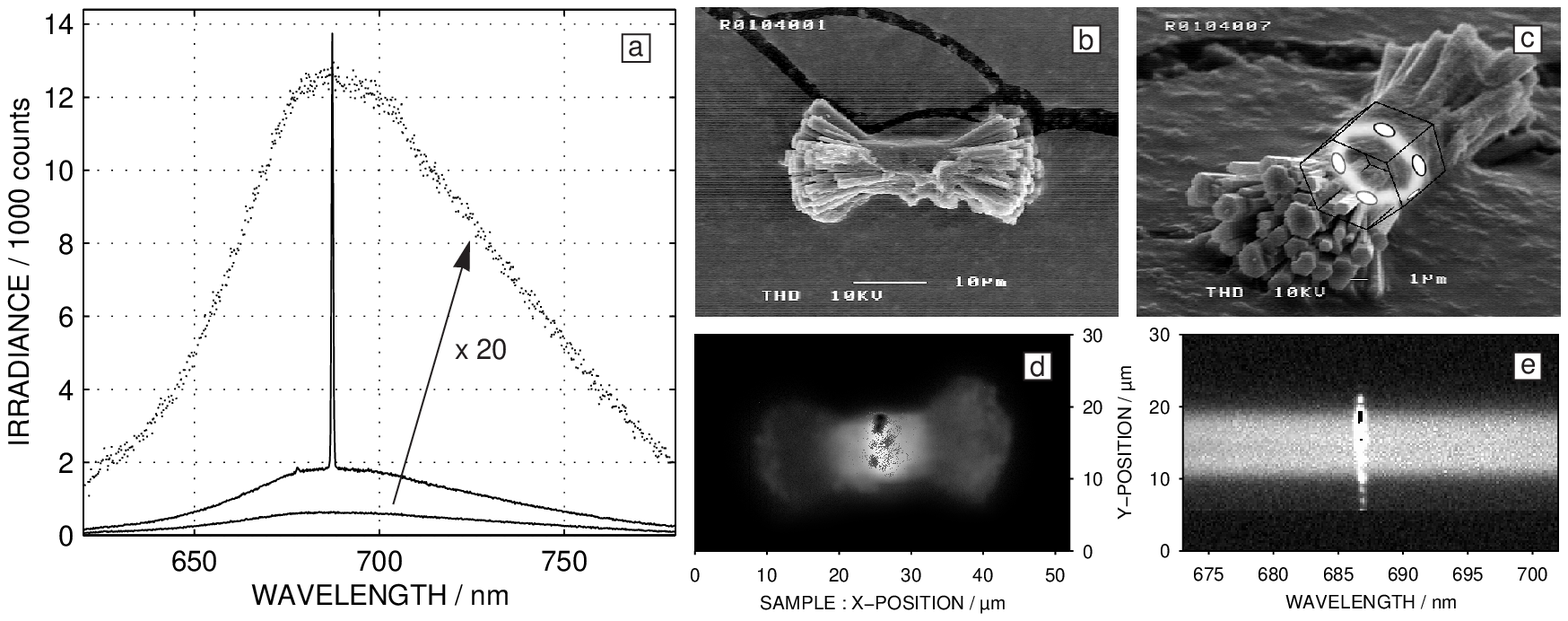} \vspace{5mm}
\begin{minipage}{178mm}
\protect\caption{(a) Emission spectra of the dumbbell shaped
AlPO$_4$-5--Pyridine2 compound microcrystal individual shown in
this figure below lasing threshold and above threshold. \protect\\
(b) Scanning electron microscope (SEM) image of the here discussed
AlPO sample.\protect\\
(c) SEM image of a microcrystal, outlining the hexagonal middle
section forming the optical resonator providing the feedback by
total internal reflection (whispering gallery mode).\protect\\
(d) Optical micrograph of the spacial distribution of the
luminescence emission taken at the same time as the subthreshold
spectrum shown in (a). Superimposed to the subthreshold
luminescence distribution, and digitally coded in black with
conventional image processing tools, are those spots where the
laser spike apparent in (a) is emitted. \protect\\
(e) On the spectrometer input the horizontal image magnification
is reduced with an anamorphotic-like imaging system so that the
whole crystal length can pass through the entrance slit. The
imaging properties of the spectrometer allow to spacially resolve
the light distribution along the entrance slit. The figure shows
the spectrum along the entrance slit (scale matched to the figure
at left), where the irradiance is rendered in grey levels. The
strong laser emission peak at 687~nm exceeds the dynamic range
spanned by the grey scale and is therefore coded with black
pixels. It is clearly shown that the  laser emission emerges from
regions corresponding to the black coded spots in (d). }
\label{Z-exp}
\end{minipage}
\end{figure}

%
%
The dye molecules were brought into the pores by in situ inclusion of
the organic dyes during the synthesis \cite{IHL97}, a method pioneered
by the group of Schulz-Ekloff and W\"ohrle \cite{WOH92}. The compound
was hydrothermally synthesized as described in ref.~\cite{DEM95}.  An
ice cold solution of triethylamine, phosphoric acid, Pyridine~2, and
water was added under stirring over 10~min to an
aluminumoxidehydroxide suspension prepared as described in
\cite{SAS96}.  The mixture was stirred for another 20~min, and filled
into a Teflon lined autoclave, which then was placed in an oven
preheated to 210$^\circ$C.  Usually, after short reaction times of
ca.~1~h, one obtains mostly regularly shaped hexagonal prisms, whereas
longer reaction times lead preferentially to dumbbell shaped crystals
where the prismatic center bar splits off into a fascicle--like
ensemble of diverging prismatic needles; cf.\ fig.\ \ref{Z-exp}(b).
These crystals are characterized by the regular shape  of their
prismatic central part, which with a width of about 8~$\mu$m, can act
as a laser resonator (fig.~\ref{Z-exp}(c)). The color distribution
observed in the light microscope indicates that the dye concentrates
in this central part, whereas the fascicled end parts appear
colorless, implying a low dye content. Consequently, laser action was
observed in the prismatic central region of the dumbbell shaped
crystals. Whichever the crystal shape, X-ray powder diffractograms
reveal no indication for the presence of any crystallographic phase
other than ($\rm P \frac{6}{\rm m}cc$)--AlPO$_4$-5.  After the desired
reaction time, the autoclaves are quench cooled, the crystals filtered
out, rinsed with water, and dried.  Their refluxing in ethanol did not
reveal any extraction of dye inclusions.  Syntheses with dye
concentrations between 0.1\%-mass and 0.01\%-mass were realized.

The long axis of the Pyridine~2 dye guest molecules lines up along
the channel pores of the AlPO$_4$-5 host, resulting in a composite
with strong macroscopic dichroic properties. The crystal appears
dark red or transparent depending on the direction of the
polarization vector of the incident light with respect to the
channel direction.  Furthermore, the Pyridine~2 molecule is an
electron push-pull-system carrying an electric dipole moment along
its axis. A priori, inclusion into the channel pores can occur with
the dipole moments all aligned in parallel, or with alternating
antiparallel orientation.  In the first case the dipole moments add
to a static macroscopic moment, whereas in the latter case the net
macroscopic moment vanishes.  Our investigations of single crystal
individuums revealed that with the inclusion of Pyridine-2 dyes, the
AlPO crystals acquire pyroelectrical properties, implying the
presence of a macroscopic moment. Thus, in contrast with the
random orientation of the dye molecules in liquid or polymeric laser
media, the dye molecules in the zeolite solid state system are not
only aligned but in average are oriented as well.

The samples are optically pumped with the second harmonic of a
Q-switch Nd:YAG-laser ($\lambda = 532$~nm, 10 ns pulse length, 10~Hz
rep.\ rate). The excited fluorescence light was collected with a
microscope objective (numerical aperture $N\!A=0.35$).  Behind the
sample the pump light was warded off with a 3~mm Schott RG 610 filter.
The imaging magnification of the sample on the CCD-chip of the cooled
imaging camera (Photometrics CH250 running at -20$^\circ$C) was
55$\times$.  In order to increase the light collection efficiency
without perturbing the imaging properties of the spectrometer (ORIEL
MS 257 with 600~L/mm, 400~nm blaze wavelength grating, and InstaSpec
IV detector, slit width ca.\ 150~$\mu$m), the magnification
perpendicular to the entrance slit was, when needed,  reduced to
ca.~5$\times$ with a cylindrical lens. Imaging camera and the
spectrometer were carefully aligned to collect and display the
fluorescence signals originating from the very same sample area. Pari
passu, the data acquisition time window of the spectrometer was
synchronized with the camera acquisition cycle.  In summary, we made
certain that the captured spectra could be uniquely associated in
space as well as in time with the images acquired by the camera.  With
this technique we analyzed samples in a heap, as well as isolated
single individual crystals.

\begin{figure}[t]
\begin{center}
\epsfxsize=80mm \epsffile{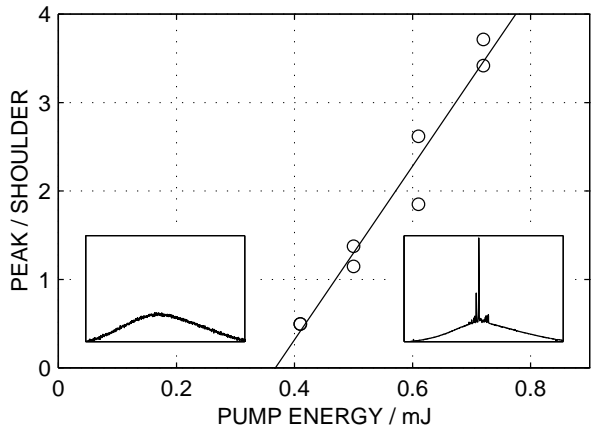}
\end{center}
\caption{ Ratio of the height of a spectral peak above shoulder to
the fluorescence shoulder height as a function of the energy of
the unfocused laser pulses ($D_{\rm Laser}=1.2$ mm) pumping an
ensemble of crystal samples. The diagram reveals the presence of a
threshold and an overproportional growth for the spectral spike
intensity. Shown in the insets are typical spectra below and above
threshold (not scaled). Note that as the laser emission (spectral
spikes) is strongly directed, not all the emitted laser light from
the samples is collected by the spectrometer aperture. }
\label{scaling}
\end{figure}

To prevent intersystem crossing processes (energy transfer to
nonluminescent, long living triplet states) quenching the
fluorescence, the system is pumped with 10~ns pulses. The oriented
inclusion of the dye molecules results in a  strongly polarized
fluorescence emission. The fluorescence distribution of the
microcrystal individuals  investigated peaks at ca.~685~nm and
exhibits a FWHM of ca.~85~nm. When the pulse energy used to pump a
number of clumped microcrystals ($>100$) is increased above a certain
threshold, we observe the sudden appearance of strong spectral spikes
around the fluorescence maximum. The spikes exhibit an instrument
limited width of ca.\ 0.6~nm and a spacing of ca.\ 4~nm. With
increasing pump energy, the spikes grow at a faster rate than the
fluorescence shoulder, as illustrated in fig.~\ref{scaling}.
%
%
%
%
The presence of a threshold energy together with the rapid
intensity increase with pump power are signatures of the
stimulated emission processes occuring in a laser. Note that in
fig.~\ref{scaling} the contribution of the collected fluorescence
light originates from a heap of microcrystals imaged to fill  the
entrance slit area, whereas the contributions to the spectral
peaks originate in small localized spots, as will be shown below.
Thus, the peak to shoulder ratio corresponding to a single
localized laser emitter is grossly underestimated in
fig.~\ref{scaling}.

Inspection of individual crystals revealed that only samples with
a dumbbell-like shape show laser emission, most of them
oscillating on several spectral lines. However, a number of
specific individuals were found to emit a single line. A typical
example is shown in fig.~\ref{Z-exp} with its emission spectrum
(a) and the corresponding SEM (b) and optical (d) micrographs. As
seen on fig.~\ref{Z-exp}(d), the strong part of the fluorescence
activity is concentrated in the prismatic central section of the
dumbbell sample within a 3~-~4~$\mu$m wide slice. This is
compatible with the distribution of the dye concentration visible
as red coloration when inspecting the sample with the light
microscope. As a consequence, this strong concentration of dyes,
and thus gain, in a thin slice can confine the oscillating field
through gain guiding. Superimposed to the below threshold
luminescence image are the localized spots, from where the strong
laser lines were emitted. Figure~\ref{Z-exp}(e) shows the spectral
distribution of the emission as a function of the image height. We
see that the strong, narrow line emission emerges from a position
which is consistent with the spots of strong emission shown in the
micrograph fig.~\ref{Z-exp}(d). As already mentioned, the
luminescence is linearly polarized parallel to the crystal axis
because of the orientation of the molecular polarizability of the
included dye molecules.  For this reason, luminescence emission,
and thus also any stimulated emission, in the direction of the
crystal axis are not possible.  Therefore, light amplification can
only occur for waves travelling in planes normal to the crystal
axis.  In fact, there exist a wide range of wavevectors parallel
to the high gain slice, which fulfill the conditions for total
internal reflection (TIR) at the prism faces.  The faces of the
prism therefore form the mirrors of an optical ring resonator in
which the necessary feedback for laser action is provided;
cf.~fig.~\ref{Z-exp}(c). The optical mode oscillating in this ring
resonator resembles a whispering gallery mode. Whispering gallery
modes were extensively discussed \cite{RAR96,DUC96,BRA89}, and as
mentioned at the beginning, lasing was observed in different
geometries. In our case the mode is confined by the prism faces
and by gain guiding in a 3~-~4~$\mu$m thick disk. We estimate the
gain induced index increase to ca. $10^{-3}$, sufficient to
confine the field \cite{HES96}. Together with the narrow width of
the gain region this leads to single frequency oscillation. In
fact, in samples in which the dye was contained in wider disks,
multiline emission was observed. The lasing threshold for single
microlasers was around $12 \times 10^{-9}$~J pump energy falling
upon the ca.~200~$\rm (\mu m)^3$ volume enclosing the lasing mode
(dye concentration ca.~0.05\%-mass). In an ideal whispering
gallery mode only minuscule amounts of diffracted light will leave
the resonator \cite{JOH92}. In the sample shown, laser light is
outcoupled at small defects at the crystal surface which disturb
the TIR; cf.~the hot spots in fig.~\ref{Z-exp}(d). The roundtrip
length for the light circulating in this ring resonator is about
$\ell\approx 25\;\mu$m, corresponding to a longitudinal mode
separation $\Delta \lambda=\lambda^2 /n \ell$ of ca.~13~nm. This
is more than three times the spectral line spacing observed in
crystals exhibiting multi--spike emission. We suspect that these
lines correspond to different morphology dependent structure
modes, anologuous to those observed in spherical lasers
\cite{TZE84}.

Although we did not systematically investigate size dependent
effects yet, we have demonstrated laser emission in composite
material microlasers with a resonator size in the regime where
size dependent effects such as reduction of laser threshold
\cite{YOK95,DEM88} start to play a role. To our knowledge the
microlasers presented here are ca.\ $3 \times$ smaller than the
smallest dye lasers realized so far \cite{KUW92}. In addition, the
dye molecules are uniformly aligned in the zeolitic solid state
host. In this system, we also observed fluorescence emission (yet
not laser emission) excited by two photon absorption of 1064 nm
laser light as well as frequency conversion of the 1064~nm
excitation to \twocolumn[
] \noindent
its 532~nm second harmonic.

This work was funded by the Deutsche Forschungsgemeinschaft DFG.

\end{document}